\documentclass[12pt]{iopart}
\usepackage{iopams}  
\begin{document}

\title[Interaction of photons with plasmas]{Interaction of photons with plasmas and liquid metals 
-----photoabsorption and scattering------}

\author{Junzo Chihara\footnote[1]{E-mail; chihara@cracker.tokai.jaeri.go.jp}}

\address{Advanced Photon Research Center, JAERI,
Kizu, Kyoto, Japan 619-0215}

\begin{abstract}
Formulas to describe the photoabsorption and the photon scattering by a plasma  
or a liquid metal are derived in a unified manner with each other. 
It is shown how the nuclear motion, the free-electron motion and 
the core-electron behaviour in each ion in the system determine 
the structure of photoabsorption and scattering in an electron-ion mixture. 
The absorption cross section in the dipole approximation consists of three terms 
which represent the absorption caused by the nuclear motion, 
the absorption owing to the free-electron motion producing optical conductivity 
or inverse Bremsstrahlung, and the absorption ascribed to the core-electron 
behaviour in each ion with the Doppler correction.
Also, the photon scattering formula provides an analysis method for 
experiments observing the ion-ion dynamical structure factor (DSF), 
the electron-electron DSF giving plasma oscillations, 
and the core-electron DSF yielding the X-ray 
Raman (Compton) scattering with a clear definition of the background scattering for each experiment, in a unified manner. A formula for anomalous X-ray scattering 
is also derived for a liquid metal. At the same time, Thomson scattering in plasma physics is discussed from this general point of view.
\end{abstract}




\section{Introduction}
For an isolated atom or the free electron system without taking account 
of the existence of bound electrons, 
the theory of photoabsorption and scattering provides a simple formula,  
as is written in the standard books \cite{AbsScat}. 
However, in a real system such as a plasma and a liquid metal, 
photons are interacting with the free electrons and the bound electrons 
in the system: we must treat the free elections and the bound electrons 
on the equal footing to investigate interactions of photons 
with matter.
In some works \cite{Grimaldi}-\cite{{Csanak}} to treat plasmas, the absorption cross section 
$\sigma(\omega)$ is separated  into 
three parts: $\sigma_{\rm bb}(\omega)$, the absorption owing to 
the bound-bound transitions in 
the ion, $\sigma_{\rm bf}(\omega)$ caused by the transitions from the bound to free state, and $\sigma_{\rm ff}(\omega)$ from the free-free transitions, 
in the formula
\begin{eqnarray}
\sigma(\omega)&=&\frac{4 \pi \omega}c \Im {\rm m} \,\alpha(\omega) \label{E:zeroAbs}\\
              &=& \sigma_{\rm bb}(\omega)+\sigma_{\rm bf}(\omega)+\sigma_{\rm ff}(\omega) 
\end{eqnarray}
with use of the atomic polarisability $\alpha(\omega)$.
The situation is not so simple as will be shown in this paper; 
photons interact with the nuclear motion through 
the carried bound electrons and the screening electrons, 
in addition to the `free' electrons and the bound electrons in the system in a coupled manner.  
Here, it should be mentioned that the free-free absorption can not be 
described by the same atomic polarisability $\alpha(\omega)$ 
to give $\sigma_{\rm bb}(\omega)$ and $\sigma_{\rm bf}(\omega)$, 
as is shown in this paper. 
This example shows some confusions in treating the photon interactions with plasmas.
It is the purpose of this paper to clarify the mechanism of photon interactions with plasmas and liquid metals.

On the other hand, the photon-scattering experiments by matter are 
focused on the observation of either the nuclear motion, 
the free-electron motion or the core-electron behaviour in the ion 
by choosing the transferred momentum ${\bf q}$ and energy $\omega$ 
suitable to the phenomena. Therefore, when the scattering by one of these motions is observed, the other motions yield the background scattering; there are many kinds of scattering named Raman, Rayleigh, Compton and Thomson combined  sometimes with the term `elastic' and/or `inelastic', for example, in a confused way.
In this paper, we present 
the scattering formula of photons by plasmas and liquid metals in a unified manner 
to observe the nuclear motion, the free-electron motion and the core-electron behaviour in the ion; this formula can clarify the relation between the observing scattering and the background scattering for each experiment,  and gives the analysis method for each experiment.

The theory of photoabsorption in atomic systems is based on 
first-order time-dependent perturbation; 
the transition probability that 
the absorption of a photon (the wavevector ${\bf q}$ and frequency $\omega$ with
 its polarisation ${\bf e_{q\lambda}}$) accompanied by a transition from a state $a$ to a state 
$b$ of the absorbing system provides the absorption 
cross section \cite{sjo64} in the form:
\begin{equation}
\fl { \sigma_{\rm a}({\bf q}\lambda,\omega)} = {4\pi^2 \over \hbar c \omega}
\left({e \over m}\right)^2\!{1 \over N}\!\sum_a p_a\!\sum_b \left|<\!\!b|\sum_{j=1}^{Z_{\rm A}N}{\bf e_{q\lambda}}{\bf p}_j
 {\rm e}^{i{\bf q}{\bf r}_j}|a\!\!>\right|^2\!\delta\left({E_a\!-\!E_b \over \hbar}\!+\!\omega\right)\,.\nonumber\\ 
\end{equation}
Here, a proper statistical averaging $p_a$ over the initial states of the absorber has been performed; $E_a$ is an eigenstate of the absorber (the nucleus-electron mixture).
This expression can be rewritten by following the manner of van Hove to derive the neutron scattering formula \cite{vH} in another form:
\begin{equation}\label{E:sigA}
\sigma_{\rm a}({\bf q}\lambda,\omega) = {4\pi^2 e^2 \over \hbar c \omega}{1 \over 2\pi N}\int_{-\infty}^\infty
 {\rm e}^{i({\bf qr}-\omega t)} a({\bf r},t)d{\bf r}dt
={4\pi^2 e^2 \over \hbar c \omega}a({\bf q},w)
\end{equation}
with
\begin{equation}
a({\bf r},t)\equiv {\bf e_{q\lambda}}\!\cdot\!\int d{\bf r}'\langle{\bf j}({\bf r}',0)
{\bf j}({\bf r}+{\bf r}',t) \rangle\cdot{\bf e_{q\lambda}}\,,
\end{equation}
in terms of the current-operator of all electrons in the system consisting of $N$ atoms with the atomic number $Z_{\rm A}$
\begin{equation}
{\bf j}({\bf r},t)\equiv \sum_{i=1}^{Z_{\rm A}N}\delta({\bf r}-{\bf r}_i(t)){{\bf p}_i(t) \over m}\,.
\end{equation}
This means that the absorption cross section can be determined by 
the current-current correlation:
\begin{eqnarray}
\mu_{\rm jj}({\bf q},\omega)&=&{1 \over N}\int {\rm e}^{i({\bf qr}-\omega t)} 
\langle{\bf j}({\bf r}',0)
 {\bf j}({\bf r}+{\bf r}',t) \rangle d{\bf r}'dt\\
&=&{1 \over N}\int_{-\infty}^\infty
\langle{\bf j}_{\bf q}^*(0)
 {\bf j}_{\bf q}(t) \rangle {\rm e}^{-i\omega t}dt
\end{eqnarray}
with
\begin{equation}\label{E:cur}
{\bf j}_{\bf q}\equiv \sum_j \frac{{\bf p}_j(t)}{m}
\exp [-i{\bf qr}_j(t)]\,.
\end{equation}
Therefore, the absorption cross section in the dipole approximation is obtained by taking the limit,
\begin{equation}\label{E:dipoleA}
 \lim_{{\bf q}\to 0}2\pi a({\bf q},\omega)
= \mu_{\rm jj}^{\rm L}(0,\omega)\equiv \mu_{q=0}^{\rm tot}(\omega)\,,
\end{equation}
where $a({\bf q},\omega)$ reduces to the longitudinal current-current correlation $\mu_{\rm jj}^{\rm L}(0,\omega)$, since the transverse current-current correlation $\mu_{\rm jj}^{\rm T}(0,\omega)$ becomes identical with the
longitudinal current-current correlation in this limit for the isotropic system.

On the other hand when an incident photon with ${\bf k}_0$, ${\bf e}_0$, $\omega_0$ is scattered by a plasma to a state ${\bf k}_1$, ${\bf e}_1$, $\omega_1$, the differential scattering cross section  is described in terms of the dynamical structure factor (DSF) $S_{\rm ee}^{\rm tot}(q,\omega)$ of all electrons involved in the plasma as free- and bound-electrons \cite{sjo64,DSF} in the form:
\begin{equation}
{d^2 \sigma \over d\Omega d\omega}=\left({d \sigma 
\over d\Omega}\right)_{\rm Th}{k_1 \over k_0}S_{\rm ee}^{\rm tot}(q,\omega) 
\label{E:totScatt}
\end{equation}
with the Thomson cross section:
\begin{equation}
\left({d \sigma \over d\Omega}\right)_{\rm Th}
\equiv \left( {e^2 \over mc^2}\right)^2({\bf e}_0\cdot{\bf e}_1)^2\,.
\end{equation}
Here, the total DSF is defined by
\begin{equation}
S_{\rm ee}^{\rm tot}(q,\omega)\equiv {1 \over 2\pi N}\int {\rm e}^{i({\bf qr}-\omega t)} 
G_{\rm ee}^{\rm tot}({\bf r},t)d{\bf r}dt\,,
\end{equation}
in terms of the time-dependent density-density correlation function:
\begin{equation}
G_{\rm ee}^{\rm tot}({\bf r},t)\equiv \int d{\bf r}'
\langle\rho_{\rm e}({\bf r}',0)\rho_{\rm e}({\bf r}+{\bf r}',t)\rangle
\end{equation}
with
\begin{equation}
\rho_{\rm e}({\bf r,t})\equiv \sum_{j=1}^{Z_{\rm A}N} \delta ({\bf r}-{\bf r}_j(t))\,.
\end{equation}

The above expressions for the absorption cross section (\ref{E:sigA}) and 
the scattering cross section (\ref{E:totScatt}) are written for all electrons, 
which are contained as the bound electrons or the free electrons coupled 
with $N$ nuclei with the atomic number $Z_{\rm A}$ in the system. 
These expressions are only formal, and do not give any information about 
the mechanism of the photoabsorption and scattering in a real system. 
Under the assumption that a plasma or a liquid metal clearly consists of 
ions with the charge $Z_{\rm I}$ 
and free electrons, we give the physical structure to the absorption cross section 
from (\ref{E:sigA}) in Sec.~\ref{PhotoAbs}, and to the scattering cross section from (\ref{E:totScatt}) in Sec.~\ref{Scatt}. The last section is devoted to conclusion.

\section{Photoabsorption}\label{PhotoAbs}
We can obtain the absorption cross section 
in the dipole approximation \eref{E:dipoleA} from the dynamical structure factor 
$S_{\rm ee}^{\rm tot}(q,\omega)$ of the total electrons  by noting the following 
relation\cite{Kubo, Hansen}:
\begin{eqnarray}\label{E:jjDSF}
\mu_{q}^{\rm tot}(\omega)&\equiv &{1 \over N}\int_{-\infty}^\infty
\langle{\bf j}_{\bf q}^*(0)
 {\bf j}_{\bf q}(t) \rangle_{_{\rm L}} {\rm e}^{-i\omega t}dt\nonumber\\
&=&2\pi {\omega^2 \over q^2}S_{\rm ee}^{\rm tot}(q,\omega)\,,
\end{eqnarray}
where the total electrons are all electrons contained in the system as the 
core electrons around each nucleus and the the free electrons.
From the definition of the DSF
\begin{equation}
S_{\rm ee}^{\rm tot}(q,\omega)\equiv {1 \over 2\pi N}
\int_{-\infty}^\infty I^{\rm tot}(q,t) {\rm e}^{i\omega t}dt\,,
\end{equation}
in terms of the intermediate scattering function:
\begin{equation}
I^{\rm tot}(q,t) \equiv  \langle \rho_e({\bf q},t)\rho_e^\ast({\bf q}, 0)
 \rangle\,,
\end{equation}
we can determine the total DSF by the total electron density $\rho_e({\bf q},t)$ from $I^{\rm tot}(q,t)$. 
Here, note that the total electron density can be split into the core-electron 
part $\rho_c({\bf q},t)$ and the free-electron part $\rho_f({\bf q},t)$:
\begin{eqnarray}
\rho_{\rm e}({\bf q},t)&\equiv &\sum_{k=1}^{Z_{\rm A}N}
 \exp[i{\bf q}\cdot{\bf r}_k(t)]\\
& = &\rho_{\rm c}({\bf q},t)+\rho_{\rm f}({\bf q},t)\,.
\end{eqnarray}
Furthermore, the core electrons are considered to be distributed among nuclei with positions ${\bf R}_\alpha$:
\begin{eqnarray}
\rho_{\rm c}({\bf q},t)&\equiv &\sum_{\alpha=1}^N\sum_{j=1}^{Z_{\bf B}}
\exp[i{\bf q}\cdot{\bf r}_{j\alpha}(t)]
=\sum_{\alpha=1}^N\sum_{j=1}^{Z_{\bf B}}
\exp[i{\bf q}\cdot({\bf r}'_{j\alpha}(t)+{\bf R}_\alpha(t))]\\
&=&\sum_{\alpha=1}^N\left(\sum_{j=1}^{Z_{\bf B}}
\exp[i{\bf q}\cdot{\bf r}'_{j\alpha}(t)]\right)
\exp[i{\bf q}\cdot{\bf R}_\alpha(t)]\,,
\end{eqnarray}
which can be approximated by using the form factor $f_{\rm I}(q)$ of the $Z_{\bf B}$ core-electrons  as follows
\begin{equation}\label{E:form}
\rho_{\rm c}({\bf q},t)\simeq  \left\langle\sum_{j=1}^{Z_{\bf B}}
\exp[i{\bf q}\cdot{\bf r}'_{j0}(t)]\right\rangle \sum_{\alpha=1}^N\exp[i{\bf q}\cdot{\bf R}_\alpha(t)]
\equiv f_{\rm I}(q)\rho_{\rm I}({\bf q},t)\,.
\end{equation}
With use of this approximation, we obtain the intermediate scattering function 
\begin{eqnarray}
I^{\rm tot}(q,t) &\equiv & \langle [\rho_{\rm c}({\bf q},t)+\rho_{\rm f}({\bf q},t)][\rho^\ast_{\rm c}({\bf q}, 0)+\rho^\ast_{\rm f}({\bf q},0)]
 \rangle\\
&=& \langle \rho_{\rm c}({\bf q},t)\rho_{\rm c}^\ast({\bf q}, 0)
 \rangle+\langle \rho_{\rm f}({\bf q},t)\rho_{\rm c}^\ast({\bf q}, 0)
 \rangle \nonumber\\
&+&\langle \rho_{\rm c}({\bf q},t)\rho_{\rm f}^\ast({\bf q}, 0)
 \rangle+\langle \rho_{\rm f}({\bf q},t)\rho_{\rm f}^\ast({\bf q}, 0)
 \rangle \\
& = &N\mid f_{\rm I}(q)\mid^2F_{\rm II}(q,t)
+N[Z_{\rm B}F^{\rm ce}(q,t)-\mid f_{\rm I}(q)\mid^2]F_{\rm s}(q,t)
\nonumber \\
&&+NZ_{\rm I}F_{\rm ee}(q,t)+2N\sqrt{Z_{\rm I}}f_{\rm I}(q)F_{\rm eI}(q,t)\,. \label{E:timeC}
\end{eqnarray}
In the above, several correlations among free electrons and ions are defined as follows:
\begin{equation}
F_{\rm II}(q,t)\equiv \sum_{\alpha, \beta}\langle\exp[i{\bf q}({\bf R}_\alpha(t)-{\bf R}_\beta(0))]\rangle/N\,,
\end{equation}

\begin{equation}
F_{\rm s}(q,t)\equiv \langle\exp[i{\bf q}({\bf R}_\alpha(t)-{\bf R}_\alpha(0))]\rangle\,,
\end{equation}

\begin{equation}
F_{\rm ee}(q,t)\equiv \langle\rho_{\rm f}(q,t)\rho^\ast_{\rm f}(q,0)\rangle/(Z_{\rm I}N)\,,
\end{equation}
 
\begin{equation}
F_{\rm eI}(q,t)\equiv \langle\rho_{\rm I}(q,t)\rho^\ast_{\rm f}(q,0)\rangle/(\sqrt{Z_{\rm I}}N)\,,
\end{equation}
and the form factor in (\ref{E:form}) is defined for the core electrons in the ion in a plasma by
\begin{equation}
f_{\rm I}(q)\equiv \langle\rho_{\rm B}({\bf q},t)
\rangle
\end{equation}
with the bound electrons density around $\alpha$-nucleus
\begin{equation}
\rho_{\rm B}({\bf q},t)\equiv \sum_{j=1}^{Z_{\rm B}}\exp[i{\bf q}\cdot{\bf r}'_{j\alpha}(t)]=\sum_{j=1}^{Z_{\rm B}}\exp[i{\bf q}\cdot{\bf r}'_{j0}(t)]\,.
\end{equation}
which becomes identical with the bound-electron density of any nucleus ($\alpha=0$).
Also, the electron-electron correlation between the core electrons in the ion 
is defined by
\begin{equation}
Z_{\rm B}F^{\rm ce}({\bf q},t)\equiv \langle\rho_{\rm B}({\bf q},t)
\rho^\ast_{\rm B}({\bf q},0) \rangle\,.
\end{equation}
From equation (\ref{E:timeC}), we can represent the total electron DSF 
in terms of the ion-ion DSF $S_{\rm II}(q,\omega)$, the electron-ion DSF $S_{\rm eI}(q,\omega)$ and the electron-electron DSF $S_{\rm ee}(q,\omega)$ in the form:
\begin{eqnarray}\label{E:See1}
S_{\rm ee}^{\rm tot}(q,\omega)&\equiv &{1 \over 2\pi N}
\int_{-\infty}^\infty I^{\rm tot}(q,t) {\rm e}^{i\omega t}dt \nonumber\\
&=&\mid f_{\rm I}(q)\mid^2S_{\rm II}(q,\omega)+2\sqrt{Z_{\rm I}}f_{\rm I}(q)S_{\rm eI}(q,\omega)+Z_{\rm I}S_{\rm ee}(q,\omega) \nonumber\\
 &&+Z_{\rm B}\int {\tilde S}^{\rm ce}(q,\omega-\omega')S_{\rm s}(q,\omega')d\omega'\,.
\end{eqnarray}
Also, the DSF between the core electrons in each ion is defined as 
\begin{eqnarray}
{\tilde S}^{\rm ce}(q,\omega)&\equiv& S^{\rm ce}(q,\omega)
 -\mid f_{\rm I}(q)\mid^2\delta(\omega)/Z_{\rm B} \label{E:bDSF2}\\
&=&{1 \over 2\pi Z_{\rm B}}\int_{-\infty}^\infty \langle\delta\rho_{\rm B}(q,t)\delta\rho^\ast_{\rm B}(q,0)\rangle{\rm e}^{i\omega t}dt \label{E:bDSF}
\end{eqnarray}
with use of the density deviation $\delta\rho_{\rm B}(q,t)$ from its average $\langle\rho_{\rm B}(q,t) \rangle$ 
\begin{equation}
\delta\rho_{\rm B}(q,t)\equiv \rho_{\rm B}(q,t)-
\langle\rho_{\rm B}(q,t) \rangle\,.
\end{equation}

With help of the the following relations derived in the previous work \cite{Chihara}
\begin{equation}\label{E:Sfee}
S_{\rm ee}(q,\omega) = {\mid \rho(q)\mid^2\over Z_{\rm I}}S_{\rm II}(q,\omega)+S^0_{\rm ee}(q,\omega)\,,
\end{equation}
\begin{equation}
S_{\rm eI}(q,\omega) = {\rho(q) \over \sqrt{Z_{\rm I}} }S_{\rm II}(q,\omega)\,,
\end{equation}
the DSF (\ref{E:See1}) of the total electrons is written as 
\begin{eqnarray}
S_{\rm ee}^{\rm tot}(q,\omega)&=&\mid f_{\rm I}(q)\!+\!\rho(q)\!\mid^2
S_{\rm II}(q,\omega)+Z_{\rm I}S^0_{\rm ee}(q,\omega) \nonumber\\
 &&+Z_{\rm B}\int {\tilde S}^{\rm ce}(q,\omega-\omega')S_{\rm s}(q,\omega')d\omega'\,.
\end{eqnarray}
Here, $\rho(q)$ represents the Fourier transforms of the electron cloud $\rho(r)$ forming a neutral pseudoatom as screening of ion
\begin{equation}
\rho(q)\equiv n_0^{\rm e}C_{\rm eI}(q)\chi_q^0
/\{1+n_0^{\rm e}\beta v_{\rm ee}(q)[1-G(q)]\chi_q^0\}\,,
\end{equation}
which is exactly prescribed by the use of the electron-ion direct correlation function $C_{\rm eI}(q)$ and the local-field correction (LFC) $G(q)$ \cite{Chihara}.
Also, $S^0_{\rm ee}(q,\omega)$ is the same form to the DSF in the jellium model 
except the the dynamical LFC $G(q,\omega)$ should be determined in an electron-ion mixture not in the jellium model, as is written in the form:
\begin{equation}\label{E:feDSF}
S^0_{\rm ee}(q,\omega)\equiv -\frac 1\pi \frac 1{n_0^{\rm e}v_{\rm ee}(q)}
{\hbar\beta \over 1-\exp(-\beta\hbar\omega)}\Im{\rm m}\left(\frac 1{\tilde{\epsilon}(q,\omega)}\right)
\end{equation}
with
\begin{equation}
\tilde{\epsilon}(q,\omega) \equiv 1+n_0^{\rm e}\beta v_{\rm ee}(q)
\frac{\chi_q^0(\omega)}
{1-n_0^{\rm e} \beta v_{\rm ee}(q)G(q,\omega)\chi_q^0(\omega)}\,,
\end{equation}
which is also an exact (but formal) expression of the `free' electron DSF.
To take account of the absorption due to the motion of nuclei, 
we must add the nucleus charge current in (\ref{E:cur}) as $e{\bf j}=e({\bf j}_{\rm e}-Z_{\rm A}{\bf J}_{\rm N})$  with 
${\bf j}_{\rm N}\equiv \sum_{\alpha}\frac{{\bf P}_{\alpha}}{M}\delta({\bf r}-{\bf R}_{\alpha})$. This contribution can be taken by calculating the following charge correlation 
\begin{equation}
e^2I^{{\rm tot}\!+\!{\rm nuc}}(q,t)\equiv e^2\langle [\rho_{\rm e}(q,t)-Z_{\rm A}\rho_{\rm N}(q,t)]
[\rho^\ast_{\rm e}(q,0)-Z_{\rm A}\rho^\ast_{\rm N}(q,0)]\rangle
\end{equation}
with
$\rho_{\rm N}(q,t)=\sum_{\alpha}\exp[i{\bf q}{\bf R}_{\alpha}(t)]$.
Thus, we obtain the final expression to derive the photoabsorption cross section
\begin{eqnarray}\label{E:TpN}
I_{\rm ee}^{\rm tot+nuc}(q,\omega)/N&=&\mid f_{\rm I}(q)\!+\!\rho(q)\!-\!Z_{\rm A}\!\mid^2
S_{\rm II}(q,\omega)+Z_{\rm I}S^0_{\rm ee}(q,\omega)\nonumber\\
 &&+Z_{\rm B}\int {\tilde S}^{\rm ce}(q,\omega-\omega')S_{\rm s}(q,\omega')d\omega'\,,
\end{eqnarray}
which with the aid of (\ref{E:jjDSF}) leads finally to the expression of the absorption cross section
\begin{eqnarray}\label{E:totabs}
\mu_q^{\rm tol}(\omega)&=&\mid f_{\rm I}(q)\!+\!\rho(q)\!-\!Z_{\rm A}\!\mid^2
\mu^{\rm nuc}_q(\omega)+Z_{\rm I}\mu^{\rm fe}_q(\omega)\nonumber\\
 &&+Z_{\rm B}\int \mu^{\rm ce}_q(\omega-\omega')S_{\rm s}(q,\omega')d\omega'\,.
\end{eqnarray}

The absorption cross section $\mu^{\rm ce}_q(\omega)$ owing to the core electrons involved the third term in (\ref{E:totabs}) provides the standard expression for the photoabsorption cross section in terms of the dipole-dipole correlation, the atomic polarisability $\hat{\alpha}(\omega)$ or 
the longitudinal conductivity $\hat{\sigma}_{\rm L}^{\rm ce}(\omega)$,
\begin{eqnarray}\label{E:absorp}
Z_{\rm B}e^2\mu^{\rm ce}_{_{q=0}}(\omega)&\equiv &\lim_{q\to 0} \int_{-\infty}^\infty
\langle e{\bf j}^{\rm ce*}_{\bf q}(0)\cdot
 e{\bf j}^{\rm ce}_{\bf q}(t) \rangle_{_{\rm L}} {\rm e}^{-i\omega t}dt \\
&=&\frac{\omega^2}{3} \int_{-\infty}^\infty
\langle {\bf d}(0)\cdot
 {\bf d}(t) \rangle {\rm e}^{-i\omega t}dt \\
&=& {2\hbar\omega^2 \over 1-\exp(-\beta\hbar\omega)}
 \Im {\rm m}\,\, \hat{\alpha}^{\rm ce}(\omega)\\
&=& {2\hbar\omega \over 1-\exp(-\beta\hbar\omega)}
 \Re{\rm e} \,\, \hat{\sigma}_{\rm L}^{\rm ce}(\omega)
\end{eqnarray}
with the definition of the atomic polarisability by the density-density response function of the core electrons in the ion:
\begin{equation}\label{E:polar}
\hat{\alpha}^{\rm ce}(\omega)\equiv -e^2\int\int z\chi^{\rm ce}({\bf r},{\bf r}';\omega)z' d{\bf r}d{\bf r}' \,.
\end{equation}
These expressions can be obtained from (\ref{E:jjDSF}) and the general relation 
\cite{Kubo} among the DSF $S_{\rm ee}^{\rm tot}(q,\omega)$, the density-density response function $\chi_q^{\rm tot}[\omega]$
and the longitudinal conductivity $\sigma_{\rm L}^{\rm tot}[q,\omega]$ as follows [see Appendix A]:
\begin{eqnarray}
S_{\rm ee}^{\rm tot}(q,\omega)&\equiv &{1 \over 2\pi N}\int_{-\infty}
^\infty\langle{ {\rho}({\bf q},t)\rho}^\ast({\bf q},0) \rangle {\rm e}^{i\omega t}dt\\
&=& -{\hbar \over 1-\exp(-\beta\hbar\omega)}
 {1 \over \pi}\Im{\rm m} \chi_q^{\rm tot}[\omega]\\
&=& {\hbar \over 1-\exp(-\beta\hbar\omega)}
 {q^2 \over \pi\omega }\Re{\rm e}\, \sigma_{\rm L}^{\rm tot}[q,\omega]\,. \label{E:cond}
\end{eqnarray}
In this way, the absorption cross section (\ref{E:sigA}) owing to the core electrons contained in each ion 
in a plasma can be written in the dipole approximation in the several forms:
\numparts
\begin{eqnarray}
\sigma^{\rm ce}_a(0, \omega)&=& \frac{4\pi \omega}{c} 
{1 \over {1-\exp(-\beta\hbar\omega)}}\Im {\rm m}\,\, \hat{\alpha}^{\rm ce}(\omega)\label{E:ionP}\\
&=& \frac{4\pi}{c} 
{1 \over {1-\exp(-\beta\hbar\omega)}}\Re {\rm e}\,\, \hat{\sigma}_{\rm L}^{\rm ce}(\omega) \label{E:ionC}\\
&=&4\pi^2\frac{e^2}{\hbar c}\omega\, \frac 1{2\pi}\!\int {\langle{\bf d}(0)\cdot{\bf d}(t)\rangle \over 3 e^2}{\rm e}^{-i\omega t}dt\\
&=&4\pi^2\frac{e^2}{\hbar c}\omega\, \sum_{ab}p_a\mid \langle b|z^{\rm ce}|a\rangle \mid^2 
\delta(\frac{E_b-E_a}\hbar -\omega) \label{E:OscSt}
\end{eqnarray}
\endnumparts
with $z^{\rm ce}\equiv \sum_{j=1}^{Z_{\rm B}}z_j$ for incident photons polarised in z-direction. 
In some works \cite{Grimaldi}-\cite{Csanak} treating high temperature plasmas, the factor $[1-\exp(-\beta\hbar\omega)]^{-1}$ in (\ref{E:ionP}) is omitted; this omission leads to the definition of the {\it net absorption} cross section \cite{Grimaldi}.

In the third term, the self-part of the ionic DSF for the high-frequency region \cite{Hansen}:
\begin{equation}
S_{\rm s}(q,\omega)=\frac{1}{2\pi}\frac{1}{v_{\rm T}q}
\exp\left[-\frac{1}{2}\left(\frac{\omega}{v_{\rm T}q}\right)^2\right]\,,
\end{equation}
gives rise to the Doppler correction to the atomic photoabsorption;
when a photon with a momentum $\hbar q_{0}\equiv \hbar \mid \omega_0-\omega_1\mid/c$ is absorbed associated with the transition between two bound levels with 
a difference frequency $\mid\omega_0-\omega_1 \mid$, the third term becomes   
\begin{equation}
\fl \int\mu_{q_0}^{\rm ce}(\omega-\omega')S_{\rm s}(q_0,\omega')d\omega'\approx \int\mu^{\rm ce}_{q=0}(\omega-\omega')
\frac{1}{\sqrt{\pi}\omega_{\rm D}}\exp\left[ -\left(\frac{\omega'}
{\omega_{\rm D}}\right)^2\right]d\omega'\,,
\end{equation}
where the Doppler width $\omega_{\rm D}$ is given by
$\omega_{\rm D}\equiv |\omega_0-\omega_1|\sqrt{2}v_{\rm T}/c$ with $v_{\rm T}^2\equiv kT/m$.

Since the absorption cross section $\mu_{q=0}^{\rm ce}(\omega)$ is derived for the nucleus-electron mixture, this can be written in the form to treat the Stark effect \cite{Dufty}:
\begin{equation}\label{E:Stark}
\fl Z_{\bf B}e^2\mu^{\rm ce}_{q=0}=\frac{\omega^2}{3} \int_{-\infty}^\infty
\langle {\bf d}(0)\cdot
 {\bf d}(t) \rangle {\rm e}^{-i\omega t}dt
=\frac{\omega^2}{3} \!\int{\rm e}^{-i\omega t}dt \int d\mbox{\boldmath $\epsilon$}q(\mbox{\boldmath $\epsilon$})
\langle {\bf d}(0)\cdot
 {\bf d}(t) \rangle_{\mbox{\boldmath $\epsilon$}}\,,
\end{equation}
using the microfield distribution
$q(\mbox{\boldmath $\epsilon$})=\langle \delta(\mbox{\boldmath $\epsilon$}-{\bf E}_{\rm i})\rangle\,.$
This expression (\ref{E:Stark}) gives a formula to treat the Stark broadening in a plasma \cite{Griem}.

The second term of (\ref{E:totabs}) represents the absorption owning to the 
free electrons:
\begin{equation}\label{E:freeAbs}
Z_{\rm I}e^2\mu_{_{q=0}}^{\rm fe}(\omega)={2\hbar\omega Z_{\rm I} \over 1-\exp(-\beta\hbar\omega)}\Re{\rm e} \hat{\sigma}_{\rm L}^{\rm fe}(\omega)\,,
\end{equation}
which can be approximated by the Drude model using the frequency-dependent collision frequency $\nu(\omega)$:
\begin{equation}
4\pi \sigma_{\rm L}^{\rm fe}(\omega)\equiv 4\pi Z_{\rm I} n_{\rm I}\hat{\sigma_{\rm L}}^{\rm fe}(\omega)\simeq \omega^2_{\rm p}/[\nu(\omega)-i\omega]\,.
\end{equation}
Thus, absorption coefficient $\kappa(\omega)=4\pi \Re{\rm e} \sigma_{\rm L}(\omega)/[ n(\omega)c]$ owing to the free electrons is written as
\begin{eqnarray}\label{E:Drude}
4\pi \Re{\rm e} \sigma_{\rm L}^{\rm fe}(\omega)=\left\{
  \begin{array}{ll}
\displaystyle{\sigma_0 \over [\omega/\nu(0)]^2+1}&
\quad \mbox{for $\omega \lesssim \omega_{\rm p}$}\\
\displaystyle{\left(\frac{\omega_{\rm p}}{\omega}\right)^2\nu(\omega)}&
\quad \mbox{for $\omega\gg \omega_{\rm p}$}
  \end{array}
  \right.\;
\end{eqnarray}
with $\sigma_0=\omega_{\rm p}^2/[4\pi\nu(0)]$ denoting the dc conductivity.
The upper case of (\ref{E:Drude}) provides the optical conductivity as was 
observed in the case of liquid Na \cite{Inagaki}, and the lower case yields 
the inverse Bremsstrahlung \cite{CR}:
\begin{equation}
k(\omega)
=\left(\frac{\omega_{\rm p}}{\omega}\right)^2
\frac{\nu(\omega)}{n(\omega)c}=\frac{\omega_{\rm p}^2
\nu(\omega)}{c\omega\sqrt{\omega^2-\omega_{\rm p}^2}}
\end{equation}
with the refractive index $n(\omega)$ [see Appendix A].

The first term of the absorption cross section (\ref{E:totabs}) 
represents the absorption owing to the nuclear motion, 
which becomes zero in this approximation, 
since $f_{\rm I}(0)=Z_{\rm B}$ and $\rho(0)=Z_{\rm I}$; 
the nucleus with the charge $Z_{\rm A}$ is perfectly neutralised 
by the bound electrons $f_{\rm I}(r)$ and the free electron cloud 
$\rho(r)$. It should be noticed that some part $\rho(q)$ of 
the free-electron contribution to the absorption is involved 
in the nuclear motion due to the relation (\ref{E:Sfee}), 
which means that the electron cloud $\rho(r)$ is attached 
to each nucleus together with the bound electrons $f_{\rm I}(r)$, 
and the other part contributes to the free-electron absorption 
as is given by (\ref{E:freeAbs}).

\section{Photon scattering}\label{Scatt}
As we have proven in the previous section, the photon scattering can be described by the 
DSF $S_{\rm ee}^{\rm tot}(q,\omega)$ of all electrons in the system:
\begin{eqnarray}\label{E:totDSF}
I(q,\omega)/N&=&S_{\rm ee}^{\rm tot}(q,\omega) \nonumber \\
&=&\mid f_{\rm I}(q)\!+\!\rho(q)\!\mid^2
S_{\rm II}(q,\omega)+Z_{\rm I}S^0_{\rm ee}(q,\omega) \nonumber\\
 &&+Z_{\rm B}\int {\tilde S}^{\rm ce}(q,\omega-\omega')S_{\rm s}(q,\omega')d\omega'\,.
\end{eqnarray}
This expression contains three dynamic structure factors: the ion-ion DSF $S_{\rm II}(q,\omega)$, which is observed usually by the thermal neutron scattering experiment, the core-electron DSF 
${\tilde S}^{\rm ce}(q,\omega)$ defined by (\ref{E:bDSF}) and the `free' electron DSF 
$S^0_{\rm ee}(q,\omega)$ defined by (\ref{E:feDSF}).

For the purpose of presenting the inelastic 
X-ray scattering for an incident photon with energy 
$\hbar \omega_0 \gg I$ (the ionisation energy), equation (\ref{E:totDSF}) is rewritten in the form to focus on $S_{\rm II}(q,\omega)$
\begin{eqnarray}
I(q,\omega)/N&=&\mid f_{\rm I}(q)\!+\!\rho(q)\!\mid^2
S_{\rm II}(q,\omega)+Z_{\rm I}S^0_{\rm ee}(q,\omega)+Z_{\rm B}{S}_{\rm inc}^{\rm I}(q,\omega)
\end{eqnarray}
with 
\begin{equation}
Z_{\rm B}{S}_{\rm inc}^{\rm I}(q,\omega)\equiv Z_{\rm B}\int {\tilde S}^{\rm ce}(q,\omega-\omega')S_{\rm s}(q,\omega')d\omega'\simeq Z_{\rm B}\tilde{S}^{\rm ce}(q,0)\,.
\end{equation}
On the basis of this equation, the dynamic structure factors $S_{\rm II}(q,\omega)$
in liquid metals are observed by Sinn \etal \cite{Burkel} by means of inelastic X-ray scattering. In this experiment, 
the second and third terms are regarded as yielding the incoherent scattering 
from the free and bound electrons.

The X-ray diffraction from a liquid metal or a plasma is described by the static 
structure factor $S_{\rm ee}^{\rm tot}(q)$:
\begin{eqnarray}\label{E:elScat}
S_{\rm ee}^{\rm tot}(q)&=&\int S_{\rm ee}^{\rm tot}(q,\omega)d\omega \nonumber\\
&=&\mid f_{\rm I}(q)+\rho(q)\mid^2S_{\rm II}(q)+Z_{\rm I}S_{\rm ee}^0(q)
+Z_{\rm B} S_{\rm inc}^{\rm I}(q)\,,
\end{eqnarray}
which is an extention of the usual formula of X-ray diffraction \cite{James} 
to the case of the metallic system \cite{Chihara}.
Here, $Z_{\rm B} S_{\rm inc}^{\rm I}(q)$ denotes the incoherent Compton scattering produced by the bound electrons:
\begin{eqnarray}
Z_{\rm B}S_{\rm inc}^{\rm I}(q)&=&Z_{\rm B} {\tilde S}^{\rm ce}(q)
 \equiv \langle\delta\rho_{\rm B}(q)\delta\rho^\ast_{\rm B}(q)\rangle\\
&\equiv& Z_{\rm B}S^{\rm ce}(q)-|f_{\rm I}(q)|^2 
\approx Z_{\rm B} -\sum_{jk}|f_{jk}(q)|^2\,,
\end{eqnarray}
which can be evaluated approximately by the formula given by James \cite{James}
in terms of $f_{lm}(q)\equiv \langle l|\exp[i{\bf q}{\bf r}]|m\rangle$, 
and $S_{\rm ee}^0(q)$ is the `free' electron structure factor.
For a liquid metal or a plasma, the incoherent Compton scattering $Z_{\rm A}S_{\rm inc}^{\rm A}(q)$ in the system consisting of neutral atoms is replaced by 
\begin{eqnarray}
Z_{\rm A}S_{\rm inc}^{\rm A}(q)&\approx &Z_{\rm A} -\sum_{jk}|f_{jk}(q)|^2\\
&\Rightarrow& Z_{\rm B} S_{\rm inc}^{\rm I}(q) + Z_{\rm I}S_{\rm ee}^0(q)\,, 
\end{eqnarray}
in conjunction with the replacement of the atomic form factor 
$f_{\rm A}(q)\Rightarrow f_{\rm I}(q)+\rho(q)$.
Since $f_{\rm A}(q)\simeq f_{\rm I}(q)+\rho(q)$ \cite{Kam,Anta}, the X-ray diffraction experiment on 
a liquid metal can be analysed by using the atomic form factor $f_{\rm A}(q)$ 
as if it is a non-metallic system.
Moreover, on the basis of (\ref{E:elScat}) Nardi \cite{Nardi} has proposed to use X-ray scattering for a diagnostic of dense strongly coupled plasmas; 
the possibility of obtaining information on the electron-ion temperature relaxation time as well as a temperature diagnostic is shown in addition to an equation of state diagnostic.

The `free' electron DSF $S_{\rm ee}^0(q,\omega)$ in (\ref{E:totDSF}) becomes 
the electron DSF $S^{\rm jell}_{\rm ee}(q,\omega)$ in the jellium model if the dynamical LFC $G(q,\omega)$ 
in (\ref{E:feDSF}) is approximated by the dynamical LFC $G^{\rm jell}(q,\omega)$ of the jellium model. 
In this sense, the inelastic X-ray scattering experiments \cite{Eisen,SNM,VP} 
for $\omega \sim \omega_{\rm p}$ provide the DSF of the electron gas in a metal. However, it should be remembered that $S_{\rm ee}^0(q,\omega)$ is only a part of the electron DSF given by (\ref{E:Sfee}). For large energy transfer $\omega \gg \omega_{\rm p}$, this term produces Compton scattering due to the free electrons \cite{Comp} 

Inelastic X-ray scattering by the core electrons in the ion \cite{Suzuki,MO,Bush} are described by the core-electron DSF ${\tilde S}^{\rm ce}(q,\omega)$ involved in (\ref{E:totDSF}).
This inelastic X-ray scattering is discriminated between Raman and Compton scattering according to 
the incoming photon energy compared with the ionisation energy.
It should be noticed that the Raman scattering cross section is closely related 
to the photoabsorption cross section, 
since the DSF ${\tilde S}^{\rm ce}(q,\omega)$ has the relation to 
the longitudinal current-current correlation $\mu_q^{\rm ce}(\omega)$ 
giving the absorption cross section as (\ref{E:jjDSF}) shows.
In the inelastic X-ray experiments, where the transfer energy $\omega$ is chosen to be suitable for 
the determination of ${\tilde S}^{\rm ce}(q,\omega)$ or $S_{\rm ee}^0(q,\omega)$
in (\ref{E:totDSF}), the first term of (\ref{E:totDSF}) provides elastic 
scattering named Rayleigh scattering \cite{Kissel}, since the ion-ion DSF can be treated 
as $S_{\rm II}(q,\omega)\sim S_{\rm II}(q)\delta(\omega)$ for this $\omega$-range.

Note that the inelastic scattering formula (\ref{E:totDSF}) of X-ray is only 
applicable to the non-resonant case ($\omega_0 \gg I$) where the incident X-ray energy $\omega_0$ is far from the ionisation energy $I$. When we treat anomalous X-ray scattering ($\omega_0 \lesssim I$) or light (and electromagnetic wave) scattering to observe the ion-ion DSF $S_{\rm II}(q,\omega)$, we must take into accout 
the second-order contribution of the ${\bf p\cdot A}$ term to scattering 
in addition to the ${\bf A}^2$ term, which alters the first term in (\ref{E:totDSF}) representing 
the scattering from the nuclear motion as
\begin{equation}\label{E:anom}
I^{\rm nuc}(q,\omega)/N\equiv \mid f_{\rm I}(q)\!+\!\rho(q)\!-\frac{m}{e^2}\omega_0^2\hat{\alpha}^{\rm ce}(\omega_0)-Z_{\rm B}\mid^2
S_{\rm II}(q,\omega)\,.
\end{equation}
Here, $\hat{\alpha}^{\rm ce}(\omega_0)$ denotes the atomic polarisability defined by (\ref{E:polar}) (see Appendix B). 
Thus, anomalous X-ray diffraction from the metallic system is described by the formula
\begin{equation}\label{E:anom2}
\fl I(q)/N\equiv \mid f_{\rm I}(q)\!+\!\rho(q)\!-
\frac{m}{e^2}\omega_0^2\hat{\alpha}^{\rm ce}(\omega_0)-
Z_{\rm B}\mid^2S_{\rm II}(q)+Z_{\rm I}S^0(q)+Z_{\rm B}
S_{\rm inc}^{\rm I}(q)\,, \label{E:AXD}
\end{equation}
which reduces to the well known formula to describe anomalous X-ray scattering \cite{James} for non-metallic systems where $\rho(q)=0$ and $Z_{\rm I}=0$. 
It is important to remember that the atomic polarisability $\hat{\alpha}^{\rm ce}(\omega_0)$ in (\ref{E:AXD}) should be determined for an ion in the presence of surrounding ions and electrons, instead of an isolated ion as is used in the analysis of anomalous X-ray diffraction experiment, and that we must take account of the presence of 
the Fermi surface of free electrons in the edge calculation, since an electron 
excited by a photon can be moved only to the above of the Fermi surface at zero temperature.  

On the other hand, in the case of light (electromagnetic wave) scattering, 
(\ref{E:anom2}) can be written as
\begin{equation}
I^{\rm nuc}(q,\omega)/N = \mid Z_{\rm I}\!-\!\frac{m}{e^2}\omega_0^2
\hat{\alpha}^{\rm ce}(0)\mid^2 S_{\rm II}(q,\omega)\,,
\end{equation}
since in this case we can approximate $f_{\rm I}(q)\!+\!\rho(q)-Z_{\rm B}\simeq Z_{\rm I}$ because of the wavevector being nearly zero (${\bf q}\sim 0$).
This expression reduces to the usual formula \cite{Gelbart} of optical Raman  
(inelastic Rayleigh) scattering for non-metallic fluids ($Z_{\rm I}=0$), 
which represents the Rayleigh line and the Brilliouin lines as elastic and inelastic scattering, respectively \cite{Hansen}.

In plasma physics, it is customary to call scattering from the free electrons 
Thomson scattering \cite{Kunze}-\cite{Myatt}, which is described by the free electron DSF:
\begin{equation}
S_{\rm ee}(q,\omega)\equiv \frac 1 {2\pi Z_{\rm I}N} \int^\infty_{-\infty} \langle\rho_{\rm f}(q,t)\rho^\ast_{\rm f}(q,0)\rangle \exp (i\omega t)dt\,. 
\end{equation}
In plasma diagnostics by means of light scattering, Thomson scattering is considered to give the ion feature in addition to the electron feature, since the bunches of electron which are a shield on each ion reflects the ion motion, and the free-electron DSF is divided into 
the electron-feature and the ion-feature parts: 
$S_{\rm ee}(q,\omega) \equiv S_{\rm ee}^{\rm e}(q,\omega) 
+S^{\rm I}_{\rm ee}(q,\omega)$, respectively.

The free-electron DSF is determined from the fluctuation-dissipation theory as
\begin{equation}\label{E:eeFluct}
S_{\rm ee}(q,\omega)= -\int^\infty_{-\infty} {\hbar \beta \over 1-\exp(-\beta\hbar\omega)}
 {1 \over \pi}\Im{\rm m} \chi_q^{\rm fe}[\omega]
\end{equation}
with the use of the free-electron density response function
\begin{equation}\label{E:feDR}
\chi^{\rm fe}_{q}[\omega]={\chi^{0e}_q[\omega] 
\over \epsilon_{\rm e}(q,\omega)}
+{\rho(q,\omega)^2 \over Z_{\rm I}}\chi^{\rm II}_q[\omega]\,.
\end{equation}
Here, 
\begin{eqnarray}
\epsilon_{\rm e}(q,\omega) & \equiv &1+n_0^e\beta v_{\rm ee}(q)
[1-G(q,\omega)]\chi^{0e}_q[\omega]\,,\\
\rho(q,\omega)&\equiv &n_0^eC_{\rm eI}(q,\omega)]\chi^{0e}_q[\omega]/\epsilon_{\rm e}(q,\omega)\,.
\end{eqnarray}
In the above expressions, the free-electron density response function is exactly (but formally) represented using the dynamical electron-ion direct correlation function $C_{\rm eI}(q,\omega)$, and $\chi^{\rm II}_q[\omega]$ is the ion-ion density response function \cite{Chihara}.
Also, the electron cloud $\rho(q,\omega)$ surrounding each ion is approximated by 
static one $\rho(q)\equiv \rho(q,0)$ in the expression (\ref{E:feDR}), 
since the electron motion is very rapid compared to the ion motion owing to the large mass difference between them. Then, there results from (\ref{E:eeFluct})
 \begin{equation}\label{E:Sfee2}
S_{\rm ee}(q,\omega) = {\mid \rho(q)\mid^2\over Z_{\rm I}}S_{\rm II}(q,\omega)+S^0_{\rm ee}(q,\omega)\,,
\end{equation}
which provides the ion feature by the first term involving the ion-ion DSF $S_{\rm II}(q,\omega)$ and the electron feature by 
the second term representing the `free' electron DSF $S^0_{\rm ee}(q,\omega)$ in (\ref{E:Sfee2}), which is already shown in (\ref{E:Sfee}) 
to derive the absorption cross section in the previous section.
It should be noted that this division of the free-electron density response 
function into the ion and electron features as (\ref{E:feDR}) is different from the usual division \cite{Kunze}-\cite{Sheffield} used for Thomson scattering in plasma physics, and is more natural than it. 

From a fundamental point of view, it is not proper to treat light scattering by plasmas only using the free-electron DSF $S_{\rm ee}(q,\omega)$, since photons are scattered also by the core-electrons coupled with the nuclear motion in a partially ionized plasma.
By taking account of the core-electron contribution to light scattering, `Thomson' scattering is described by
\begin{equation}\label{E:eThomson}
\fl I^{\rm Thomson}(q,\omega)/N\equiv \mid f_{\rm I}(q)\!+\!\rho(q)\!-\frac{m}{e^2}\omega_0^2\hat{\alpha}^{\rm ce}(0)-Z_{\rm B}\mid^2
S_{\rm II}(q,\omega)+Z_{\rm I}S^0_{\rm ee}(q,\omega)\,,
\end{equation}
and the third term ${\tilde S}^{\rm ce}(q,\omega)\simeq {\tilde S}^{\rm ce}(q,0)$ of (\ref{E:totDSF}) yields the background scattering for this experiment.
When the wavevector $q$ is small and the atomic polarisability is negligible compared with $Z_{\rm I}$, this expression reduces to 
\begin{equation}
I^{\rm Thomson}(q,\omega)/N \simeq  Z_{\rm I}^2
S_{\rm II}(q,\omega)+Z_{\rm I}S^0_{\rm ee}(q,\omega)\,, \label{E:eThomson0}
\end{equation}
which becomes identical with (\ref{E:Sfee2}) due to $\rho(0)=Z_{\rm I}$, 
and makes assurance of the conventional experimental analysis for the small q-region.
In this formula, the total scattering cross section is determined by
\begin{eqnarray}\label{E:eThomsonT}
\fl I^{\rm Thomson}(q)/N
&=& \mid f_{\rm I}(q)\!+\!\rho(q)\!-\frac{m}{e^2}\omega_0^2\hat{\alpha}^{\rm ce}(0)-Z_{\rm B}\mid^2
S_{\rm II}(q)+Z_{\rm I}S^0_{\rm ee}(q)\\
&\simeq&Z_{\rm I}^2S_{\rm II}(q)+Z_{\rm I}S^0_{\rm ee}(q)\,,
\end{eqnarray}
from which the ion-feature and electron-feature parts at the zero wavevector are provided in the forms: $I_{ion}^{\rm Thomson}(0)/N=\mid Z_{\rm I}-\frac{m}{e^2}\omega_0^2\hat{\alpha}^{\rm ce}(0)\mid^2n_0^{\rm I}\kappa_T/\beta$ ($\kappa_T$: the isothermal compressibility) and $I_{\rm ele}^{\rm Thomson}(0)/N=0$, respectively,  
while the large q behaviours are like $\lim_{q \to \infty}I_{\rm ion}^{\rm Thomson}(q)/N=\mid Z_{\rm B}+\frac{m}{e^2}\omega_0^2\hat{\alpha}^{\rm ce}(0)\mid^2$ and 
$\lim_{q \to \infty}I_{\rm ele}^{\rm Thomson}(q)/N=Z_{\rm I}$. These results show  different behaviours from the 
ion and electron features of the usual definition,
 even in the weakly coupled classical-gas limit
  $n_0^{\rm I}\kappa_T/\beta=1/(Z_{\rm I}+1)$.

In addition to light scattering, it should be mentioned that inelastic X-ray scattering may be used for plasma 
diagnostics on the basis of (\ref{E:totDSF}); the electron feature ${\tilde S}^{\rm ce}(q,\omega)$ may 
be easily observed. On the other hand, 
it will be difficult to measure the ion feature $S_{\rm II}(q,\omega)$, of which observation requires a high resolution experiment ($\Delta\omega\sim 10$ meV), although the ion-ion DSF's of liquid metals have been observed by Sinn \etal \cite{Burkel} using inelastic X-ray scattering. 

\section{Conclusion}
On the basis of the dipole approximation, we have derived the photoabsorption 
cross section (\ref{E:totabs}) for a plasma or a liquid metal, 
which represents photoabsorption caused by the nuclear motion. 
the free electron motion and the core-electron behaviour in the ion; 
Also, each term of (\ref{E:totabs}) can be described, if necessary, in any form of the dipole-dipole 
correlation,  the polarisability, the conductivity or 
oscillator strengths, as is written for the case of the core-electron photo-absorption in (\ref{E:ionP})-(\ref{E:OscSt}). It should be noticed that some part $\rho(q)$ of free-electrons contributes to the photoabsorption due to the nuclear motion as the screening charge, while the other part constitutes the so-called free-free absorption; this absorption does not come from the atomic polarisation $\alpha^{\rm ce}(\omega)$ of (\ref{E:ionP}) as was frequently recognized in the form $\alpha_{\rm ff}^{\rm ce}(\omega)$.

The absorption cross section in the dipole approximation is related to 
the photon scattering cross section as the relation (\ref{E:jjDSF}) indicates.
In the derivation of the absorption cross section, the form factor approximation (\ref{E:form}) plays an important role to derive the expression of the total DSF. It is important to remember that this approximation is well established one in the X-ray diffraction experiments. Also, the `metallic' form factor $f_{\rm I}(q)+\rho(q)$ should be thought as an established approximation, since the X-ray diffraction of all liquid metals 
is analysed with use of the atomic form factor $f_{\rm A}(q)$, which is almost identical with the metallic form factor $f_{\rm I}(q)+\rho(q)$. This experimental fact makes assurance of the photoabsorption expression (\ref{E:totabs}) through the relation (\ref{E:jjDSF}).

On the other hand, the photon scattering formulas, (\ref{E:totDSF}) and (\ref{E:anom}), provide 
a wide view point of photon scattering experiments; we can see 
the whole mechanism of photon scattering in a single formula, which can be used to observe any motion in the system containing 
the nuclear motion, the free-electron motion, or the core-electron 
behaviour in the ion. Since the structure of the background scattering for each experiment is clearly defined, a combination of several kinds of experiments 
provides reliable data each other.

In usual analysis of the anomalous X-ray scattering experiment, 
the anomalous scattering factor 
$\frac{m}{e^2}\omega_0^2\hat{\alpha}^{\rm ce}(\omega_0)\!-\!
Z_{\rm B}$ is taken from the result calculated for a neutral atom 
even in a metal.
It should be noted that the anomalous form factor for the metallic system is 
different from the non-metallic system; $f_{\rm A}(q)$ is replaced 
by $f_{\rm I}(q)+\rho(q)$ and the atomic polarisability 
$\hat{\alpha}^{\rm ce}(\omega_0)$ should be calculated in a metallic state 
taking account of effects of the surrounding ions and electrons 
in addition to the presence of the Fermi surface.

The so-called Thomson scattering in plasma physics is nothing but 
`light' scattering from a plasma; a light scattering formula (\ref{E:anom}) for a plasma 
derived in the present work have proved that `Thomson' scattering is 
described only by the free-electron DSF for the small q-region. 
Also, we proposed a more natural division of Thomson scattering into 
the ion feature and the electron feature as is given by (\ref{E:Sfee2}).

The most fundamental assumption to derive the photoabsorption and 
scattering formula is that a liquid metal or a plasma can be considered 
as an electron-ion mixture. In a liquid metal, this model has 
no problem especially for a simple metal. 
On the other hand, in a plasma the ion in this theory is only the average 
ion; there are many kinds of charge states in general. 
Therefore, in order to compare the result of this model with experiments, 
we need further refinement such as a combined use of the Saha equation to 
determine the charge population \cite{CUK}: this problem is remained for a 
future work.
\section*{Appendix A}
\appendix
\setcounter{section}{1}
The photoabsorption cross section owing to the core electrons in the ion 
can be represented by the core-electron DSF as
\begin{eqnarray}
\mu_{q}^{\rm ce}(\omega)&\equiv &{1 \over Z_{\rm B}}\int_{-\infty}^\infty
\langle{\bf j}_{\bf q}^{{\rm ce}*}(0)
 {\bf j}_{\bf q}^{\rm ce}(t) \rangle_{_{\rm L}} {\rm e}^{-i\omega t}dt\\
&=&2\pi {\omega^2 \over q^2}S_{\rm ee}^{\rm ce}(q,\omega)=
2\pi {\omega^2 \over q^2}{\tilde S}_{\rm ee}^{\rm ce}(q,\omega)\,, \label{E:SandS}
\end{eqnarray}
where equation (\ref{E:SandS}) results from the fact that $S_{\rm ee}^{\rm ce}(q,\omega)$ is different from ${\tilde S}_{\rm ee}^{\rm ce}(q,\omega)$ only by the term 
$\delta(\omega)$ as is defined in (\ref{E:bDSF2})
Therefore, the zero-$q$ limit of the core-electron DSF divided by $q^2$:
\begin{equation}
\frac{ {\tilde S}^{\rm ce}(q,\omega) }{q^2} \equiv {1 \over 2\pi Z_{\rm B}}
\int_{-\infty}^\infty \frac{ {\tilde I}^{\rm ce}(q,t)}{q^2} {\rm e}^{-i\omega t}dt\,,
\end{equation}
can be calculated by taking this limit of
\begin{eqnarray}
\fl \lim_{{\bf q}\to 0}{{\tilde I}^{\rm ce}(q,t) \over q^2} &\equiv&
\lim_{{\bf q}\to 0}\frac1{q^2}\int\int d{\bf r}d{\bf r}'
{\rm e}^{-i{\bf qr}}\langle \delta\rho_{\rm B}({\bf r}, 0)
\delta\rho_{\rm B}({\bf r}',t)\rangle{\rm e}^{i{\bf qr}'}\\
&=&\int\int d{\bf r}d{\bf r}'z\langle \delta\rho_{\rm B}({\bf r}, 0)\delta\rho_{\rm B}({\bf r}',t)\rangle z'\\
&=&\langle z(0)z(t)\rangle = \frac 1 3 \langle {\bf r}(0)\cdot{\bf r}(t)\rangle\,,
\end{eqnarray}
which gives rise to the dipole-dipole correlation
\begin{eqnarray}
\lim_{{\bf q}\to 0}{e^2{\tilde S}^{\rm ce}(q,\omega) \over q^2} 
&=&{1 \over 2\pi Z_{\rm B}}
\int_{-\infty}^\infty \frac 1 3 \langle {\bf d}(0)\cdot{\bf d}(t)\rangle{\rm e}^{-i\omega t}dt
\end{eqnarray}
with
\begin{equation}
{\bf d}(t)=e{\bf r}(t)\equiv e{\bf r}\sum_{i\in \rm bound} \delta({\bf r}-{\bf r}_i(t))\,.
\end{equation}
In a similar way, we can obtain the atomic polarisability by taking this limit 

\begin{eqnarray}
\lim_{{\bf q}\to 0}{e^2\chi_q^{\rm ce}[\omega] \over q^2}
&\equiv &\lim_{{\bf q}\to 0}\frac1{q^2}\int\int d{\bf r}d{\bf r}'{\rm e}^{-i{\bf qr}}
\chi^{\rm ce}({\bf r},{\bf r}';\omega)
{\rm e}^{i{\bf qr}'}\\ 
&=&\frac{e^2}{3}\int\int d{\bf r}d{\bf r}'({\bf r}\cdot{\bf r}')\chi^{\rm ce}({\bf r},{\bf r}';\omega)\\
&\equiv & -{\hat \alpha}^{\rm ce}(\omega)\,.
\end{eqnarray}
Here, the density-density response function $\chi_q^{\rm ce}[\omega]$ is defined explicitly by 
\begin{eqnarray}
\chi_q^{\rm ce}[\omega]\hspace{-0cm}&\equiv &\frac1{Z_{\rm B}} \int_0^{\infty}{\rm e}^{i\omega t}
\langle \frac{1}{i\hbar} 
[\rho_q(t), \rho_q^{\ast}(0)]
\rangle dt\\
\hspace{-2cm}&=& \frac 1 {Z_{\rm B}} \sum_{ba}p_a\!\mid \langle b|\rho_{\bf q}^\ast
|a\rangle\mid^2\left\{ {1 \over \omega-\omega_{\rm ba}+i\eta}
-{1 \over \omega+\omega_{\rm ba}+i\eta}\right\}
\end{eqnarray}
with $\hbar \omega_{\rm ba}\equiv E_b-E_a$. 

On the other hand, equation (\ref{E:cond}) results from the relation between the conductivity and the polarisability;

\begin{equation}
\hat{\sigma}_{\rm L}(\omega)=-i\omega\hat{\alpha}(\omega)\,,
\end{equation}
which is the zero $q$-limit of the general relation between the longitudinal conductivity 
and the density response function \cite{Kubo}
\begin{equation}
\hat{\sigma}_{\rm L}[q,\omega]\equiv 
\frac{e^2}{N}\int_0^\infty\beta\langle{\bf j}_{\bf q}^*(0);
 {\bf j}_{\bf q}(t) \rangle_{_{\rm L}} 
{\rm e}^{i\omega t}dt=i\omega{e^2 \chi_q[\omega] \over q^2}\,.
\end{equation}

The absorption coefficient $k(\omega)$ \cite{Born} is defined by the complex refractive index 
$\hat{n}$, which is given by the dielectric constant $\hat{\epsilon}$
\begin{equation}
\hat{\epsilon}(\omega)\equiv 
1+{4\pi \sigma_{\rm L} \over \omega}i
=1+4\pi n_{\rm I}\hat{\alpha}(\omega)\,,
\end{equation}
in the form:
\begin{equation}
k(\omega)\equiv {2\omega\Im {\rm m}\,\,\hat{n} \over c}
={2\omega\Im {\rm m}\,\,\hat{\epsilon}^{1/2} \over c}\,.
\end{equation}
This definition of $k(\omega)$ provides an approximated expression for the case that $\hat{\epsilon}''\ll \hat{\epsilon}'$
:
\begin{equation}
k(\omega)\approx {\omega \hat{\epsilon}'' (\omega) \over n(\omega)c}
={4\pi \Re{\rm e} \sigma_{\rm L}(\omega)\over n(\omega)c}
={4\pi\omega n_{\rm I}\Im{\rm m} \hat{\alpha}(\omega)\over n(\omega)c}
\end{equation}
with the refractive index $n(\omega)\approx \sqrt{\Re{\rm e} \hat{\epsilon}(\omega)}$.

\section{}

Interaction of electrons with the radiation field ${\bf A}({\bf r})$ is described by the Hamiltonian
\begin{equation}
H_{\rm I}=H_1 + H_2\,,
\end{equation}
consisting two terms: the ${\bf p}\cdot{\bf A}$ term, 
\begin{equation}
\fl H_1=-{e^2 \over 2mc^2}\sum_{i}{\bf p}_i{\bf A}({\bf r}_i)
 \Rightarrow-{e \over mc}\sum_{\bf q}\left({2\pi\hbar c^2 \over V\omega_{\bf q}}\right)^{1/2}
 \{P({\bf q})a_{\bf q}+P^{\dagger}({\bf q})a_{\bf q}^{\dagger}\}\,,
\end{equation}
which is in the second quantized representation to describe the photon absorption, 
and the ${\bf A}^2$ term,
\begin{eqnarray}
\fl &H_2&\equiv\frac{e^2}{2mc^2}\sum_i A_i^2\\
&\Rightarrow&{e^2 \over 2mc^2}\sum_{{\bf kk}'}\left({2\pi\hbar c^2 \over V\sqrt{\omega_{\bf k}\omega_{\bf k'}}}\right)({\bf e}_{\bf k}\cdot{\bf e}_{\bf k'})
\left[\sum_j {\rm e}^{i({\bf k-k'})\cdot{\bf r}_j}\right][a_{\bf k}a_{{\bf k'}}^{\dagger}
+a_{\bf k'}^{\dagger}a_{\bf k}]\,, 
\end{eqnarray}
which generates the photon scattering, with 
\begin{equation}\label{E:P}
P({\bf k})\equiv \sum_j ({\bf e}_{\bf k}\cdot{\bf p}_j)
\exp[i{\bf k}\cdot{\bf r}_j]\,.
\end{equation}
Since any j-bound electron can be thought to belong to some $\alpha$-ion in the system, 
we can represent its coordinate as ${\bf r}_j={\bf R}_\alpha+{\bf r}_{j\alpha}$:
therefore, we can obtain an approximate expression for (\ref{E:P}) in the form:
\begin{equation}
P({\bf k})\approx \sum_\alpha 
\exp[i{\bf k}\cdot{\bf R}_\alpha]P^\alpha({\bf k}) \label{E:appP}
\end{equation}
with
\begin{equation}
P^\alpha({\bf k})\equiv \sum_j ({\bf e}_{\bf k}\cdot{\bf p}_{j\alpha})\exp[i{\bf k}\cdot{\bf r}_{j\alpha}]\equiv P^0({\bf k})\,.
\end{equation}
As a result of (\ref{E:appP}), we can define the transition matrix of each ion $\alpha$ from an initial state $I=\omega_0,{\bf k}_0,{\bf e}_0$ to a final state $F=\omega_1,{\bf k}_1,{\bf e}_1$ by 
\begin{eqnarray}
F^\alpha_{\rm FI}(\omega_0,\omega_1)&\equiv& \langle F_e|H_2^\alpha|I_e\rangle +\sum_N {\langle F_e|H_1^\alpha|N\rangle 
\langle N|H_1^\alpha|I_e\rangle \over E_I-E_N} \nonumber\\
&= &{\rm e}^{i({\bf k}_0-{\bf k}_1){\bf R}_\alpha}\left\{\langle F_e|\sum_i\exp[i({\bf k}_0\!-\!{\bf k}_1)\cdot{\bf r}_{i0}]|I_e\rangle ({\bf e}_0\cdot{\bf e}_1)\right. \nonumber\\
&&\left. +\sum_N {\langle F_e|H_1^{\rm ce}|N\rangle 
\langle N|H_1^{\rm ce}|I_e\rangle \over E_I-E_N}\right\}\,.
\end{eqnarray}
Therefore, the transition matrix for all ions in the system is written as
\begin{eqnarray}
F_{\rm FI}(&\omega_0,&\omega_1;\{R_\alpha\})\equiv \sum_\alpha F^\alpha_{\rm FI}(\omega_0,\omega_1)\nonumber\\
&=&\left[\sum_\alpha{\rm e}^{i({\bf k}_0-{\bf k}_1){\bf R}_\alpha}\right] \left\{
\langle F_e|\sum_i\exp[i({\bf k}_0\!-\!{\bf k}_1)\cdot{\bf r}_{i0}]|I_e\rangle ({\bf e}_0\cdot{\bf e}_1)\right. \nonumber\\
&&\ \ \ -\frac{1}{m}\sum_N \left[ {\langle F_e|P^0({\bf k}_0)|N\rangle 
\langle N|P^{0\dagger}({\bf k}_1)|I_e\rangle \over \epsilon_N-\epsilon_I+\hbar \omega_1} \right. \nonumber\\
& &\ \ \ \ \ \ \ \ \left.\left.+{\langle F_e|P^{0\dagger}({\bf k}_1)|N\rangle 
\langle N|P^0({\bf k}_0)|I_e\rangle \over \epsilon_N-i\eta-\epsilon_I-\hbar \omega_0} 
\right] \right\} \label{E:transP}\,.
\end{eqnarray}
Now, the second term in the large bracket can be written in the elastic approximation 
\cite{Cromer} in the form:
\begin{eqnarray}
\fl \hspace{1.8cm}\frac{1}{m}\sum_N &&\left[ 
{\langle I_e|P^{0\dagger}({\bf k}_0)|N\rangle 
\langle N|P^0({\bf k}_0)|I_e\rangle \over \hbar\omega_0-\hbar \omega_{N0}+i\eta}
-{\langle I_e|P^0({\bf k}_0)|N\rangle 
\langle N|P^{0\dagger}({\bf k}_0)|I_e\rangle \over \hbar\omega_0+\hbar \omega_{N0}} 
 \right]\\
&&= {\bf e}_1\cdot m\chi^{\rm ce}_{JJ}(-{\bf k}_0,\omega_0)\cdot{\bf e}_0
={\bf e}_1\cdot m\chi^{\rm ce}_{JJ}({\bf k}_0,\omega_0)\cdot{\bf e}_0\,,
\end{eqnarray}
in terms of the current-current correlation function $\chi^{\rm ce}_{JJ}({\bf k}_0,\omega_0)$ \cite{PN} for the core electrons in each ion ($\hbar \omega_{N0}\equiv \epsilon_N-\epsilon_I $).
Because of the relation \cite{PN} 
\begin{equation}\label{E:diAp}
\fl \lim_{k \to 0}m\chi^{\rm ce}_{JJ}({\bf k},\omega)=\left[\lim_{k \to 0}
m\frac{\omega^2}{k^2}\chi_k^{\rm ce}[\omega]-Z_{\rm B}\right]\!{\bf 1}
=\left[-\frac m{e^2}\omega^2{\hat \alpha}(\omega)-Z_{\rm B}\right]\!{\bf 1}\,, \label{E:diAp2}
\end{equation}
which shows that the current-current correlation 
in the zero wavevector limit can be represented by the atomic 
polarisability, 
the transition matrix (\ref{E:transP}) in the dipole approximation (\ref{E:diAp2}) can be written finally in the form:
\begin{eqnarray}
F(\omega_0;\{R_\alpha\})&=&[f_{\rm I}({\bf q})
-m\omega_0^2{\hat \alpha}(\omega_0)/e^2-Z_{\rm B}]  
\left[\sum_\alpha {\rm e}^{i{\bf q}{\bf R}_\alpha (t)}\right]({\bf e}_0\cdot{\bf e}_1)\\
&\equiv &F_{\rm ion}(q,\omega_0)\rho_{\rm I}(q,t)({\bf e}_0\cdot{\bf e}_1)
\end{eqnarray}
with ${\bf q}={\bf k}_0-{\bf k}_1$.
This indicates that, for the resonant case, the approximation $\rho_c(q,t)=f_{\rm I}(q)\rho_{\rm I}(q,t)$ of (\ref{E:form}) 
should be replaced by $\rho_c(q,t)=F_{\rm ion}(q,\omega_0)\rho_{\rm I}(q,t)$
which leads to the expression for anomalous scattering (\ref{E:anom}).

\end{document}